\documentclass[11pt,a4paper]{article} 

\usepackage{jcappub} 
\usepackage{graphicx}
\usepackage{dcolumn}
\usepackage{bm}
\usepackage[english]{babel}
\usepackage[latin1]{inputenc} 
\usepackage{graphicx}
\usepackage{epstopdf}
\usepackage{amsfonts}
\usepackage{color}
\usepackage{epsfig}
\usepackage{mathrsfs}
\usepackage{amsmath,amssymb}
\usepackage{subfigure}

\newcommand{\lto}[1]{\longrightarrow#1}

\renewcommand{\(}{\left(}
\renewcommand{\)}{\right)}
\renewcommand{\[}{\left[}
\renewcommand{\]}{\right]}

\begin{document}

\graphicspath{{figure/}}
\selectlanguage{english}

\title{Multiscale autocorrelation function: a new approach to anisotropy studies}

\author[a,b]{M. De Domenico,}
\author[b,c]{A. Insolia,}
\author[d,e,f]{H. Lyberis,}
\author[b,c]{M. Scuderi}

\affiliation[a]{Laboratorio sui Sistemi Complessi, Scuola Superiore di Catania, \\
Via Valdisavoia 9, 95123 Catania, Italy}
\affiliation[b]{Istituto Nazionale di Fisica Nucleare, Sez. di Catania, \\
Via S. Sofia 64, 95123 Catania, Italy}
\affiliation[c]{Dipartimento di Fisica e Astronomia, Universit\'a degli Studi di Catania, \\
Via S. Sofia 64, 95123 Catania, Italy}
\affiliation[d]{ IPN Orsay CNRS/IN2P3 and Universit\'e Paris Sud, \\
15 Rue Georges Clemenceau, 91406 Orsay, France}
\affiliation[e]{Universit\'e Paris VII Denis Diderot, \\
Paris 7, 75205 Paris Cedex 13, France}
\affiliation[f]{Dipartimento di Fisica Generale, Universit\'a di Torino, \\
Via Pietro Giuria 1, 10125 Torino, Italy}

\emailAdd{manlio.dedomenico@ct.infn.it}
\emailAdd{antonio.insolia@ct.infn.it}
\emailAdd{lyberis@ipno.in2p3.fr}
\emailAdd{mario.scuderi@ct.infn.it}

\date{\today}

\abstract{
We present a novel catalog-independent method, based on a scale dependent approach, to detect anisotropy signatures in the arrival direction distribution of the ultra highest energy cosmic rays (UHECR). The method provides a good discrimination power for both large and small data sets, even in presence of strong contaminating isotropic background. We present some applications to simulated data sets of events corresponding to plausible scenarios for charged particles detected in the last decades by world-wide surface detector-based observatories.
}

\arxivnumber{1001.1666}
\notoc

\maketitle

\flushbottom

\section{Introduction}

In many field involving data analysis, the search for anisotropy has played a crucial role.
Many estimators, namely correlation functions \cite{Davis-1983, Szalay-1993, Hamilton-1993}, have been proposed and widely used  to search for clustering of objects and to measure deviation from isotropy of angular distributions. These methods apply to angular coordinates of objects as well to distributions of arrival directions of events: in this work we will indifferently refer to both as \emph{arrival direction distributions of events}.

If $n$ is the number of experimental coordinates on some region $\mathcal{S}$ of a spherical surface and $r$ is the number of coordinates coming from several Monte Carlo realizations of $\mathcal{S}$, the common anisotropy analysis involves the computation of different estimators: Peebles (or natural), Davis-Peebles, Landy-Szalay and Hamilton \cite{Davis-1983, Szalay-1993, Hamilton-1993} angular correlation function, respectively defined as
\begin{eqnarray}
\omega_{1}(\Theta) &=& \frac{r(r-1)}{n(n-1)}\frac{DD(\Theta)}{RR(\Theta)}-1\\
\omega_{2}(\Theta) &=& \frac{2r}{n-1}\frac{DD(\Theta)}{DR(\Theta)}-1\\
\omega_{3}(\Theta) &=& \frac{r(r-1)}{n(n-1)}\frac{DD(\Theta)}{RR(\Theta)}-\frac{r-1}{n}\frac{DR(\Theta)}{RR(\Theta)}+1\\
\omega_{4}(\Theta) &=& \frac{4nr}{(n-1)(r-1)}\frac{DD(\Theta)\times RR(\Theta)}{DR^{2}(\Theta)}-1
\end{eqnarray}
where $DD$ is the number of pairs lying between $\Theta$ and $\Theta+\Delta \Theta$ for the experimental distribution, $RR$ is the same number calculated for Monte Carlo realizations and $DR$ is the cross-pair counts between experimental and simulated events. By definition (see Ref. \cite{Peebles-1980}) the function $\omega_{i}(\Theta)$ ($i=1,2,3,4$) is strictly related to the excess on the area $d\Omega$  of the pairs number $dN_{\text{pairs}}$ over the random background $dN_{\text{MC}}$:
\begin{eqnarray}
dN_{\text{pairs}}-dN_{\text{MC}}=\frac{1}{2}nn_{0}\omega_{i}(\Theta)d\Omega
\end{eqnarray}
where $n_{0}$ is the expected average density of events on $\mathcal{S}$ assuming an isotropic distribution.

A variant of the two-point angular correlation function is widely used for small data set of points \cite{kachelriess2005ultra, kachelriess2006clustering, cuoco2006first, cuoco2008clustering, cuoco2009global}. It is defined as
\begin{eqnarray}
w\(\Theta\)&=& \sum_{i=1}^{n}\sum_{j=i}^{i-1} \text{H}\(\Theta-\Theta_{ij}\)  
\end{eqnarray}
where H is the Heaviside function and 
\begin{eqnarray}
\label{angdist}
\Theta_{ij}=\arccos\(  \cos\theta_i \cos \theta_j  \cos (\phi_i - \phi_j)+ \sin \theta_i \sin \theta_j  \) \quad
\end{eqnarray}
is the angular distance between two directions $i$ and $j$ with coordinates $(\phi,\theta)$ on the sphere. 

Recently, new estimators have been introduced to study the anisotropy signature of sky's arrival direction distributions: the modified two-point Rayleigh \cite{ave20092pt+}, and shape-strength method derived from a principle component analysis of triplets of events \cite{Hague09}. Such a test statistics have been recently applied to both P. Auger data and to synthetic maps of events, the latter generated by sampling the Veron-Cetty \& Veron catalog \cite{VCV06} of nearby candidate active galactic nuclei (AGN) within 75 Mpc ($z\leq0.018$), showing a higher discrimination power than other estimators \cite{MelloNeto2009}.

Within the present work, we introduce a new fast and simple method for anisotropy analysis, which makes use of a multiscale approach and depends on one parameter only, namely the angular scale of the instrinsic anisotropy. The main advantage of our estimator is the possibility to analytically treat the results: the analytical approach drastically reduces computation time and makes available the possibility of applications to very large data sets of objects. We test the method on several simulated isotropic and anisotropic arrival direction distributions (mock maps) and perform an extensive analysis of its statistical features under both the null and the alternative hypotheses. However, it is worth remarking that the scope of applicability of our method is not limited to UHECR physics, and it is valid for any distribution of angular coordinates of objects.


\section{Multiscale Autocorrelation Function}\label{Scale}

Let $\mathcal{S}$ be a region of a spherical surface and let $P_{i}\(\phi,\theta\)$ ($i=1,2,...,n$) be a set of points locating $n$ arrival directions on $\mathcal{S}$, defining a \emph{sky}. The sky $\mathcal{S}$ is partitioned within a grid of $N$ equal-area (and almost-equal shape) disjoint boxes $\mathcal{B}_{k}$ ($k=1,2,...,N$) as described in Ref. \cite{stokes2004using}. Let $\Omega$ be the solid angle covered by $\mathcal{S}$, whereas each box $\mathcal{B}_{k}$ covers the solid angle 
\begin{eqnarray}
\Omega_{k} = \frac{1}{N}\int_{\theta_{\text{min}}}^{\theta_{\text{max}}}\int_{\phi_{\text{min}}}^{\phi_{\text{max}}}d\cos\theta d\phi = 2\pi(1-\cos\Theta)\nonumber
\end{eqnarray}
where $2\Theta$ is the apex angle of a cone covering the same solid angle: $N,\Theta$ and $\Omega$ are deeply related quantities that define a scale.

Let $\psi_{k}(\Theta)$ be the fraction of points in the data set falling into the box $\mathcal{B}_{k}$: the function $A(\Theta)$ that quantifies the deviation of data from an isotropic distribution at the scale $\Theta$, is chosen to be the Kullback-Leibler divergence \cite{kullback1951information, kullback1987kullback}
\begin{eqnarray}
\label{def-A}
A(\Theta) = \mathcal{D}_{\text{KL}}\( \psi(\Theta) || \overline{\psi}(\Theta)\) = \sum_{k=1}^{N} \psi_{k}(\Theta)\log\frac{\psi_{k}(\Theta)}{\overline{\psi}_{k}(\Theta)}
\end{eqnarray}
where $\overline{\psi}_{k}(\Theta)$, generally a function of the domain meshing, is the expected fraction of points isotropically distributed on $\mathcal{S}$ falling into the box $\mathcal{B}_{k}$. The Kullback-Leibler divergence is an information theoretic measure widely used in hypothesis testing and model selection criteria \cite{akaike1973information,akaike1974new,anderson2000null}, statistical mechanics \cite{plastino1997relationship, plastino1997minimum, portesi2007geometrical}, quantum mechanics \cite{fuchs1995distinguishability, reginatto1998derivation, abe1999quantum, abe2003nonadditive}, medical \cite{gersch1979automatic} and ecological \cite{burnham2001kullback} studies, to cite some of the most known. This measure quantifies the error in selecting the fraction $\overline{\psi}(\Theta)$ to approximate the fraction $\psi(\Theta)$ and it is strictly connected to maximum likelihood estimation (see Appendix \ref{app-KL}). It is straightforward to show that $A(\Theta)$ is minimum for an isotropic distribution of points, or, in general, when $\psi(\Theta)\sim \overline{\psi}(\Theta)$, i.e. if the model is correct.

If $A_{\text{data}}(\Theta)$ and $A_{\text{iso}}(\Theta)$ refer, respectively, to the data and to an isotropic realization with the same number of events, we define \emph{multiscale autocorrelation function} (MAF) the estimator
\begin{eqnarray}
\label{def-s}
s(\Theta)=\frac{\left|A_{\text{data}}(\Theta)-\left\langle A_{\text{iso}}(\Theta) \right\rangle\right|}{\sigma_{A_{\text{iso}}}(\Theta)}
\end{eqnarray}
where $\left\langle A_{\text{iso}}(\Theta) \right\rangle$ and $\sigma_{A_{\text{iso}}}(\Theta)$ are the sample mean and the sample standard deviation, respectively, estimated from several isotropic realizations of the data. If $\mathcal{H}_{0}$ denotes the null hypothesis of an underlying isotropic distribution for the data, the chance probability at the angular scale $\Theta$, properly penalized because of the scan on $\Theta$, is the probability 
\begin{eqnarray}
\label{def-p}
p(\Theta) = \text{Pr}\( s_{\text{iso}}(\Theta') \geq s_{\text{data}}(\Theta) | \mathcal{H}_{0}, \forall \Theta'\in\mathcal{P}\)
\end{eqnarray}
obtained from the fraction of null models giving a multiscale autocorrelation, at any angular scale $\Theta'$ in the parameter space $\mathcal{P}$, greater or equal than that of data at the scale $\Theta$. The null hypothesis is eventually rejected in favor of the alternative $\mathcal{H}_{1}=\lnot \mathcal{H}_{0}$ $-$ being $\lnot$ the negation operator $-$ at the angular scale $\Theta$, with probability $1-p(\Theta)$.

Under the null hypothesis $\mathcal{H}_{0}$, the estimator $s(\Theta)$ follows a half-Gaussian distribution, independently on the value of the angular scale $\Theta$ and on the number of events on $\mathcal{S}$, as it will be successively shown in the text.


\section{Dynamical counting}\label{sec-dyncount}

\begin{figure}[!t]  
\centering
  	\subfigure[\, Static counting]{ 
	  \includegraphics[width=4cm]{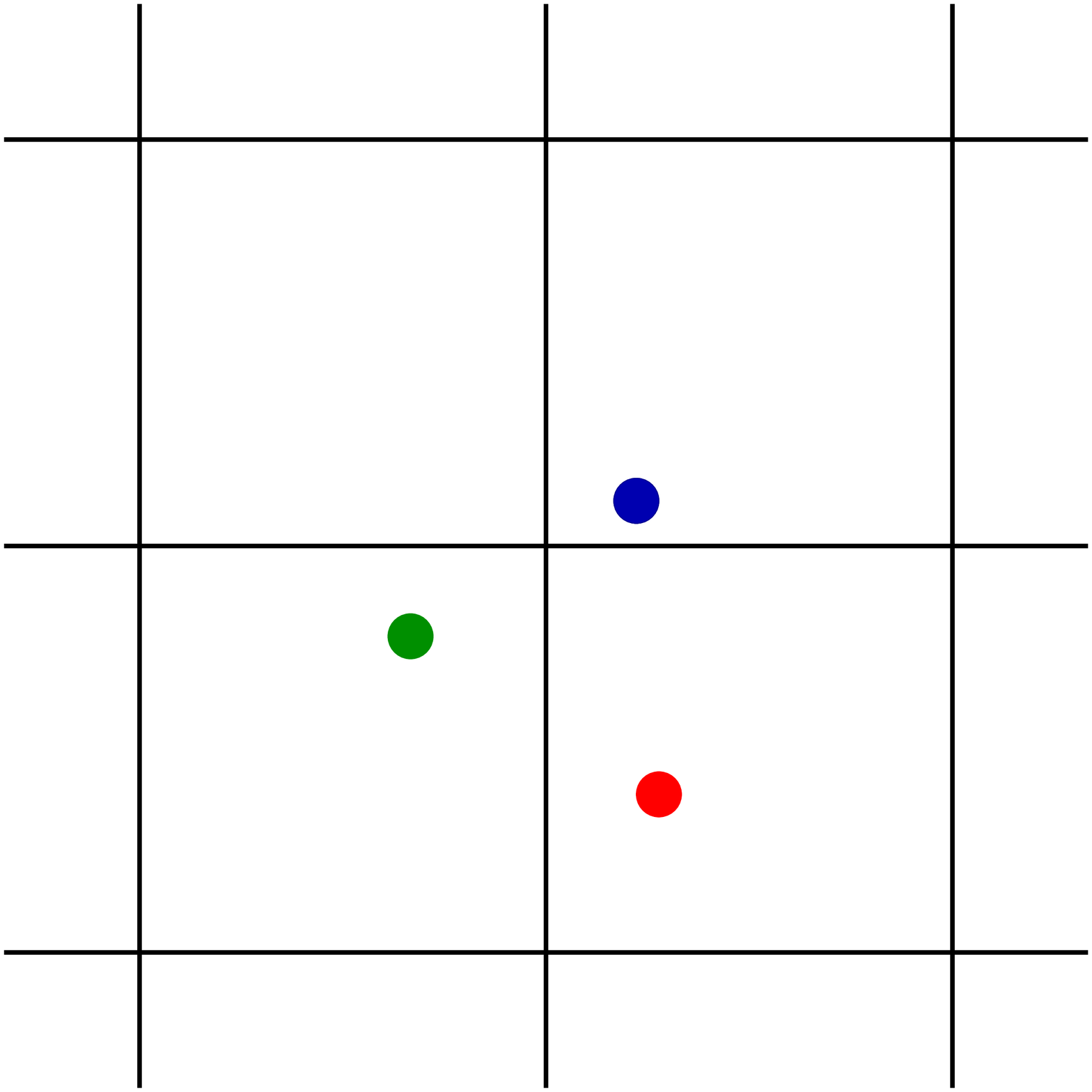}}
	\subfigure[\, Dynamical counting]{ 
	  \includegraphics[width=4cm]{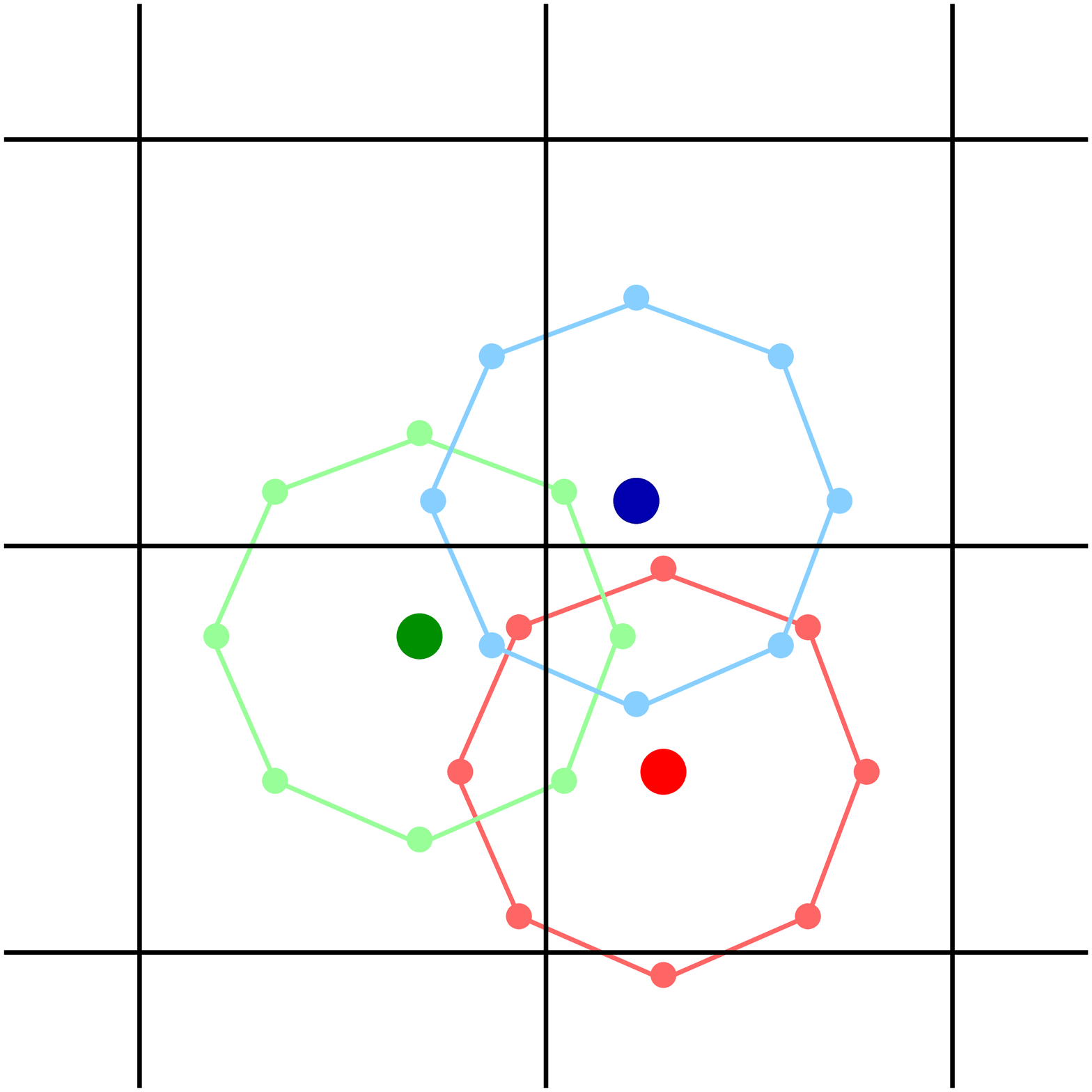}}
	  \caption{A cluster of three points falling into different boxes.}
\label{points}
\end{figure}

The simplest definition of the counting algorithm, as shortly described in Section \ref{Scale}, involves the fixed grid introduced in Ref. \cite{stokes2004using}, where each box only embodies the relative number of events falling in it.
Unfortunately, such a \emph{static counting} approach could not reveal an existing cluster. For instance, Figure\,\ref{points}a shows a typical scenario where some points of a given triplet fall into different cells. Indeed, the fixed grid may cut a cluster of points within one or more edges, causing a further loss of information at the angular scale under investigation.
To overcome this possible loss of information, we introduced a type of smoothing of the grid by applying it on the data. 

The smoothing, adopted in our study, deals with a new counting procedure for the estimation of the density $\psi_{k}(\Theta)$. However, such a density depends on the observatory's exposure, generally a function on the celestial sphere depending on both the latitude of the experiment and the maximum zenith of detection, quantifying the effective time-integrated detection area for the flux of particles from each observable sky position. The relative exposure $\omega$ is the dimensionless function corresponding to the exposure normalized to its maximum value. For a single full-time operating detector, i.e. with constant exposure in right ascension, fully efficient for particles arriving with zenith angles smaller than $\theta_{\text{max}}$, it depends on the declination $\delta$ and it is defined as \cite{sommers2001cosmic}
\begin{eqnarray}
\label{exposure}\omega(\delta)\propto \cos\phi_{0}\cos\delta\sin\alpha_{m}+\alpha_{m}\sin\phi_{0}\sin\delta,
\end{eqnarray}
where $\phi_{0}$ is the detector latitude and
\begin{eqnarray}
\alpha_{m}=\left\{
\begin{array}{ll}
0 & \xi>1\\
\pi & \xi<-1\\
\cos^{-1}\xi & \text{otherwise}
\end{array}
\right.
\end{eqnarray}
with
\begin{eqnarray}
\xi\equiv \frac{\cos\theta_{\text{max}}-\sin\phi_{0}\sin\delta}{\cos\phi_{0}\cos\delta}.\nonumber
\end{eqnarray}

Given an angular scale $\Theta$, for each point $P_{i}$ falling inside a box, we consider a set of 8 new points lying on a virtual box centered on $P_{i}$. Let $\alpha_{i,0}$ and $\delta_{i,0}$ be, respectively, the right ascension and the declination of the point $P_{i}$. We introduce the following notation:
\begin{eqnarray}
\delta_{i,\pm1} &=& \delta_{i,0} \pm\frac{\Theta}{2}\nonumber\\
\alpha_{i,\pm1} &=& \alpha_{i,0} \pm g\(\delta_{i,0}\)\nonumber\\
\alpha_{i,\pm2} &=& \alpha_{i,0} \pm g\(\delta_{i,+1}\)\nonumber\\
\alpha_{i,\pm3} &=& \alpha_{i,0} \pm g\(\delta_{i,-1}\)\nonumber
\end{eqnarray}
where $g(\cdot)$ is a function obtained from Eq. (\ref{angdist}), depending on declination, that constrains the angular distance between each of the 8 points and the original one to be $\frac{\Theta}{2}$.
Within this framework, for each $P_{i}(\alpha_{i,0},\delta_{i,0})$, we have the following 9 \emph{extended} points:
\begin{itemize}
\item The original point $P_{i}(\alpha_{i,0},\delta_{i,0})$;
\item The up $P_{i}(\alpha_{i,0},\delta_{i,+1})$ and down $P_{i}(\alpha_{i,0},\delta_{i,-1})$ points;
\item The left $P_{i}(\alpha_{i,-1},\delta_{i,0})$ and right $P_{i}(\alpha_{i,1},\delta_{i,0})$ points;
\item The up-left $P_{i}(\alpha_{i,-2},\delta_{i,+1})$ and up-right $P_{i}(\alpha_{i,2},\delta_{i,+1})$ points;
\item The down-left $P_{i}(\alpha_{i,-3},\delta_{i,-1})$ and down-right $P_{i}(\alpha_{i,3},\delta_{i,-1})$ points.
\end{itemize}
We define the function $h(\delta_{i,j})=\omega(\delta_{i,j})/\omega(\delta_{i,0})$ ($j=0,\pm1$) and introduce the weights
\begin{eqnarray}
\label{extpointweigth}
f(\delta_{i,j})=\frac{h(\delta_{i,j})}{3h(\delta_{i,-1})+3+3h(\delta_{i,+1})}
\end{eqnarray}

In other words, we weight the angular region around a given direction with the local value of the exposure. Finally, we follow the procedure described in Section \ref{Scale} by using the weighted distribution of points instead of the original one, as shown in Figure\,\ref{points}b: thus, the density function $\psi(\Theta)$ is defined as the fraction of \emph{extended} points, opportunely weighted the by function $f(\delta_{i,j})$ defined in Eq. (\ref{extpointweigth}). Our numerical studies show that such a \emph{dynamical counting} approach recovers the correct information on the amount of clustering in the data. 

\begin{figure}[!t]  
\centering
  	\subfigure[\, Clustered points]{ 
	  \includegraphics[width=4cm]{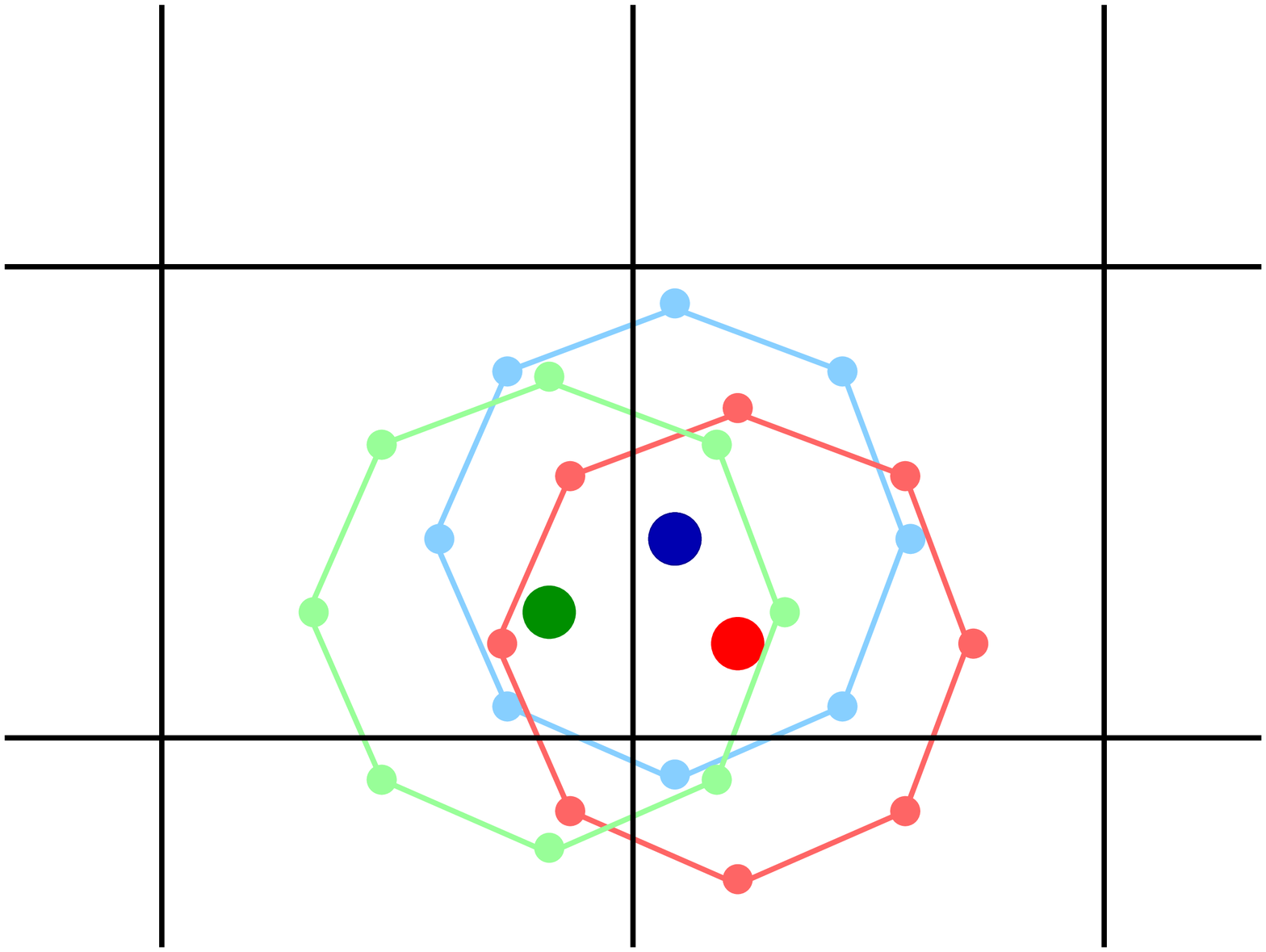}}
	\subfigure[\, Unclustered points]{ 
	  \includegraphics[width=4cm]{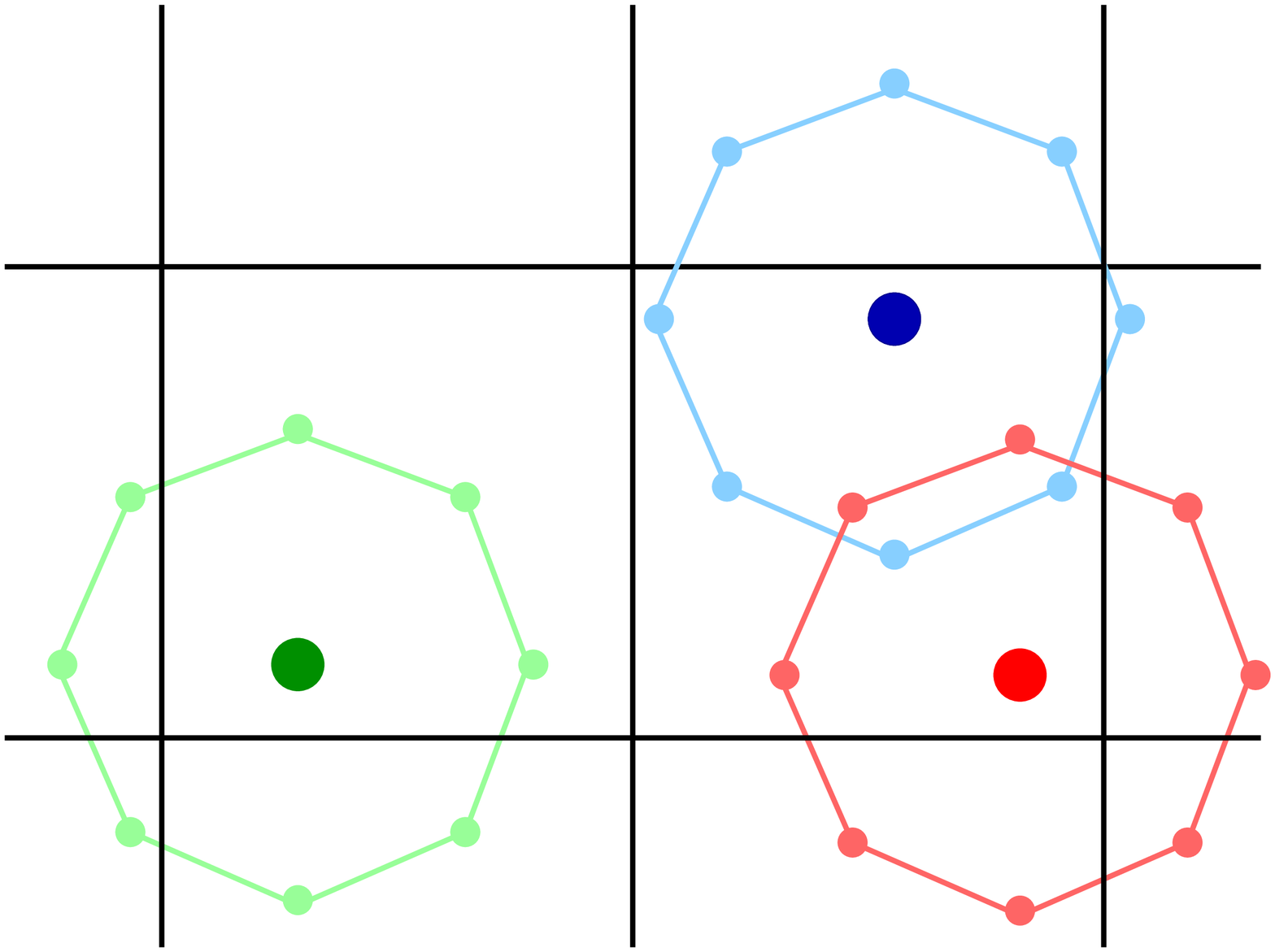}}
	  \caption{a) Three clustered points: the \emph{extended} points are mainly concentrated in two adjacent boxes. b) Three unclustered points: the \emph{extended} points are mainly distributed on the neighbor cells.}
\label{cluster}
\end{figure}

In fact, the main difference between the static and the dynamical counting lies in the value of the estimator when the procedure is applied to Monte Carlo realizations of the sky. For instance, let us consider the Figure \ref{cluster}, where we show a clustered (Figure \ref{cluster}a) and an unclustered (Figure \ref{cluster}b) set of points. The static counting is not able to recover the differences between the two configurations. Conversely, if the dynamical counting is applied, the \emph{extended} points in Figure \ref{cluster}a are concentrated in two adjacent boxes while in Figure \ref{cluster}b they are distributed on the neighbor cells. This fundamental difference is reflected in the density function, leading to two different $\psi(\Theta)$. Monte Carlo skies producing the same clustered configuration shown in Figure \ref{cluster}a, and of consequence the same weight distribution, are not frequently expected: in this case, the value of $s(\Theta)$ should be greater than that one estimated from the static method. The direct consequence of a greater value of the estimator $s(\Theta)$ is a lower chance probability and the main advantage of using the dynamical counting, instead of the static one, should be the lowest penalization of $s(\Theta)$ only if an anisotropy signal is really present.



\begin{figure}[!t]  
\centering
	  \includegraphics[width=15cm]{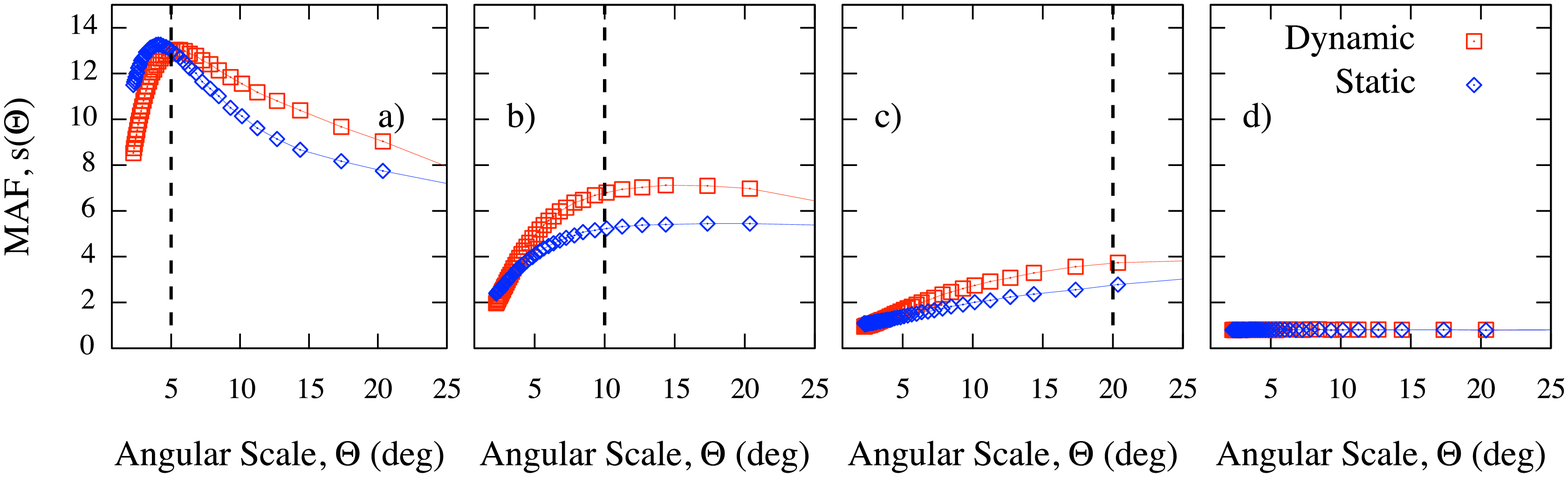}
	  \caption{MAF: average $s(\Theta)$ (solid line) estimated from $5000$ isotropic and anisotropic skies of 100 events each. In each anisotropic sky, 60\% of events are normally distributed, with dispersion $\rho$, around 10 random sources and 40\% of events are isotropically distributed. The dashed line indicates the value of the dispersion adopted to generate the corresponding mock map: a) 5, b) 10 and c) 20 degrees; d) isotropic map.}
\label{staticvsdyn}
	  \includegraphics[width=15cm]{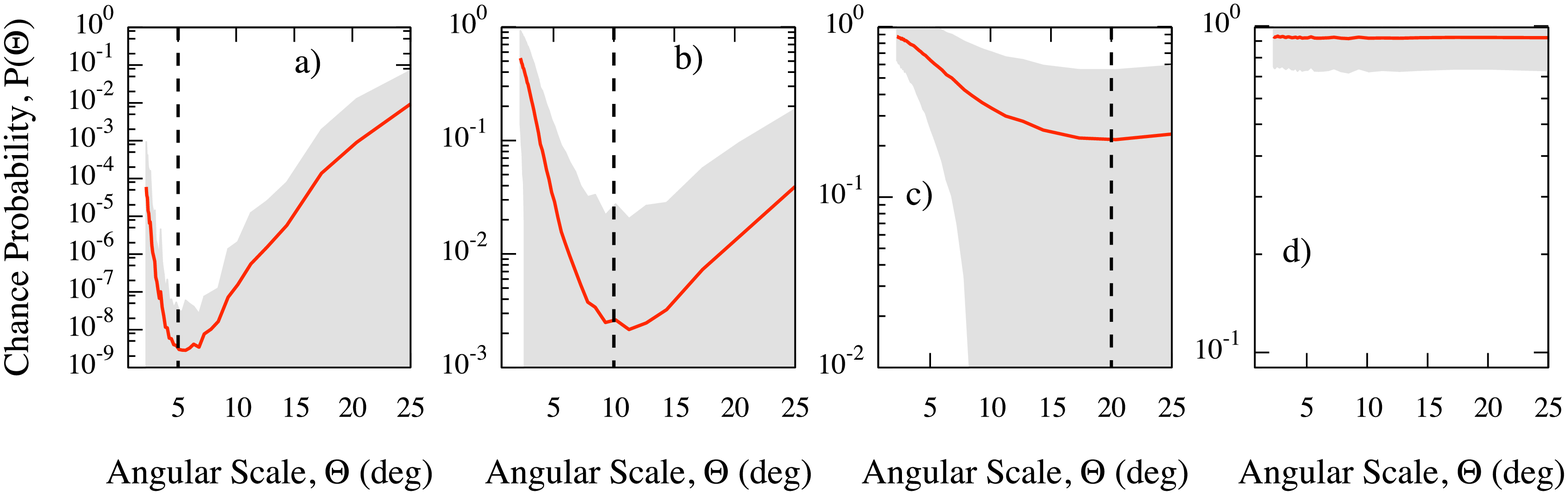}
	  \caption{MAF: average chance probability (solid line), with 68\% region around the mean value, estimated from isotropic and anisotropic skies generated as explained in Figure \ref{staticvsdyn}. Dynamical counting is used. The dashed line indicates the value of the dispersion adopted to generate the corresponding mock map: a) 5, b) 10 and c) 20 degrees; d) isotropic map.}
\label{AngScaleMock2}
\end{figure}

 In order to illustrate the importance of dynamical counting in the anisotropy signal detection, we have generated 5000 isotropic and anisotropic skies of 100 events each. In each anisotropic sky, 60\% of events are normally distributed, with dispersion $\rho$, around 10 random sources and 40\% of events are isotropically distributed. For each angular scale $\Theta$, we have estimated the average value of $s(\Theta)$ with the static and the dynamical counting, separately. Results are shown in Figure \ref{staticvsdyn} for $\rho=5^{\circ}$ (a), $\rho=10^{\circ}$ (b), $\rho=20^{\circ}$ (c) and for the isotropic map (d). As expected, the two counting methods do not show differences in the estimation of MAF in the case of isotropic skies, resulting in the same flat average value of $s(\Theta)$. Conversely, in the case of the anisotropic skies, the dynamical counting provides a greater estimation of $s(\Theta)$ than the static counting, leading to a smaller estimation of the corresponding chance probability and improving the signal-to-noise ratio. In the next section we will show how the dynamical counting is able to correctly recover the most significant clustering scale. For sake of completeness, we have generated all mock maps with a full-sky coverage and a uniform exposure.

\section{Interpretation of MAF}

Any catalog-independent method provides information about the angular scale $\Theta^{\star}$ where the significance is minimum. In the case of a simple two-point method, such an angular scale is quite difficult to interpret and topologically different configurations of events lead to the same significance. In the case of the modified two-point Rayleigh method, the estimation of the significance includes another set of parameters, independent from the angular distribution, as described in Ref. \cite{ave20092pt+}: parameters are sensitive to the orientation of the pairs and therefore to skies showing preferential directions and filamentary structure of points. It follows that $\Theta^{\star}$ is the most significant angular size for the mix of these informations, still linked to the pair configuration. In the case of the shape-strength method, the estimation of two parameters, namely the shape and strengh, is performed: both can be interpreted, respectively, in terms of size and elongation of the triangles defined by a triplet of points. It follows that all information is recovered from the configuration of triplets. 


In the specific case of MAF, the angular scale $\Theta^{\star}$, where the significance is minimum, turns to be the significative \textit{clustering scale}: it is the scale at which occurs a greater accumulation of points respect to that one occuring by chance, with no regard for a particular configuration of points, e.g. doublets or triplets. To illustrate better the clustering scale detection feature of MAF, we have generated 5000 isotropic and anisotropic skies of 100 events each, as previously described at the end of Section \ref{sec-dyncount}. Figure\,\ref{AngScaleMock2} shows the average chance probability, with 68\% region around the mean value, versus the angular scale for three values of the dispersion, namely $\rho=5^{\circ}$ (a), $\rho=10^{\circ}$ (b), $\rho=20^{\circ}$ (c), and for the isotropic map (d). As expected, chance probability is close to one and nearly flat in the case of the isotropic map, because all clustering scales are equally likely. Conversely, for all anisotropic maps, the average chance probability gets a minimum around the corresponding value of $\rho$. Thus, our estimator is able to recover the most significant clustering scale. It should be remarked that when the $20^{\circ}$ dispersion is used, the angular scale of the minimum is less obvious because of the large fluctuations due to the isotropic contamination. Finally, it is worth noticing that we have observed that the curve around the value of $\rho$ gets narrower by increasing the number of events.


\section{Statistical analysis of MAF}

In this section, we investigate the statistical features of MAF by inspecting its behavior under both the null or the alternative hypothesis. In particular, we estimate the significance $\alpha$ (or Type I error), i.e. the probability to wrongly reject the null hypothesis when it is actually true, and the power $1-\beta$ (where $\beta$ is known as Type II error), i.e. the probability to accept the alternative hypothesis when it is in fact true. In the following we will adopt the dynamical counting previously discussed.

\begin{figure}[!t]
	\centering
	  \includegraphics[width=12cm]{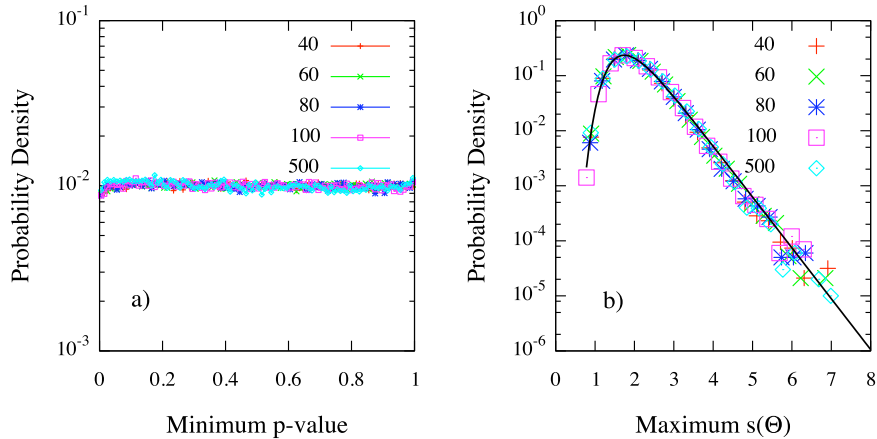}
	\caption{MAF. a) Distribution of $\tilde{p}(\Theta^{\star})$ for $n=40, 60, 80, 100$ and $500$ events. b) Distribution of $\max\{s(\Theta)\}$ for $n=40, 60, 80, 100$ and $500$ events. Solid line correspond to the least-square fit of the Gumbel density with parameters $\mu=1.743  \pm 0.002$ and $\sigma=0.470   \pm 0.002$ ($\chi^{2}/\text{ndf}=1.1\times 10^{-5}$).}
\label{smax-gumbel}
\end{figure}

\vspace{0.25truecm}\hspace{0.1truecm}\emph{Null hypothesis.} We generate isotropic maps of $10^{5}$ skies, by varying the number of events from 20 to 500: for each sky in each map, we estimate the MAF for several values of the angular scale $\Theta$. Hence, we choose the value of $\Theta=\Theta^{\star}$ where the chance probability is minimum, as the most significant clustering scale: 
\begin{eqnarray}
\tilde{p}(\Theta^{\star})=\arg\min_{\Theta}p(\Theta)\nonumber
\end{eqnarray}
properly penalized because of the scan on the parameter $\Theta$, according to the definition in Eq. (\ref{def-p}). Indipendently on the number of events in the mock map, we find an excellent flat distribution of probabilities $\tilde{p}(\Theta^{\star})$, shown in Figure \ref{smax-gumbel}a for skies of different size, as expected for analyses under the null hypothesis $\mathcal{H}_{0}$. In other words, MAF is not biased against $\mathcal{H}_{0}$, as required for good statistical estimators.

\begin{figure}[!t]
	\centering
	  \includegraphics[width=12cm]{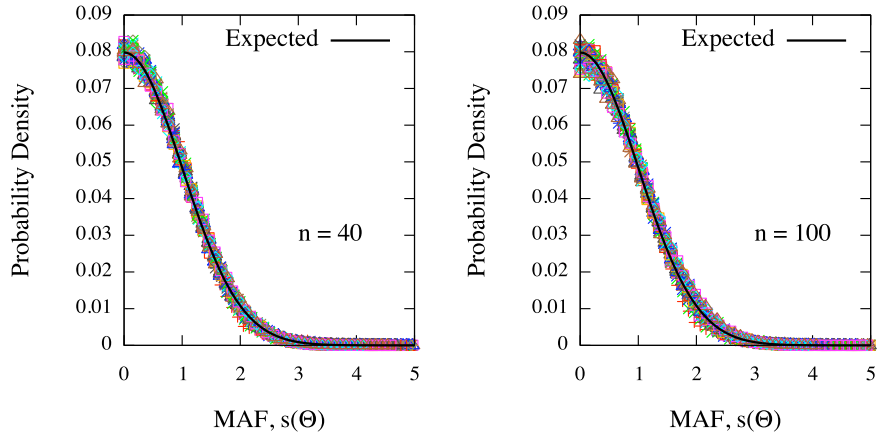}
	\caption{Distribution of the MAF estimator for $n=40$ and $n=100$ events, for angular scales $\Theta$ ranging from $2^{\circ}$ to $26^{\circ}$, separately. Solids lines correspond to the expected half-normal distribution.}
\label{sall-halfnormal}
\end{figure}

Because of the definition in Eq. (\ref{def-A}) and of the central limit theorem, a Gaussian distribution is expected for the function $A(\Theta)$, and of consequence, the half-normal distribution 
\begin{eqnarray}
\mathcal{G}_{1/2}[s(\Theta)] = \frac{2}{\sqrt{2\pi}\sigma(\Theta)}e^{-\frac{s^{2}(\Theta)}{2\sigma^{2}(\Theta)} }
\end{eqnarray}
for $\sigma(\Theta)=1$, is expected for the estimator $s(\Theta)$ defined as in Eq. (\ref{def-s}), being normalized to zero mean and unitary variance. In Figure \ref{sall-halfnormal} are shown the distributions of the MAF estimator for $n=40$ and $n=100$ events, for angular scales $\Theta$ ranging from $2^{\circ}$ to $26^{\circ}$, separately. We find an excellent agreement between the distribution for Monte Carlo realizations and the expected one. It follows that the (unpenalized) probability to obtain by chance a value of the MAF, greater or equal than a given value $s_{0}$, is just $1-\text{erf}\(\frac{s_{0}}{\sqrt{2}}\)$, being erf the standard error function, independently of the angular scale $\Theta$.

Although this important feature of the MAF estimator, generally the distribution of $s_{\text{max}}=\max\{s(\Theta)\}$ is of interest for applications, because of the required penalization due to the scan over the parameter $\Theta$. Hence, it is important to identify the distribution of the penalized probability $p(\Theta)$, if any. Intriguingly, our numerical studies show that such a distribution exists and it corresponds to one of the limiting densities in the extreme value theory (see Appendix \ref{app-Gumbel}). In particular, the cumulative density of maxima is known as the Gumbel distribution \cite{gumbel1954statistical,gumbel2004statistics}:
\begin{eqnarray}
G(x)=\exp\[ -\exp\(\frac{x-\mu}{\sigma}\) \]\nonumber
\end{eqnarray}
where $\mu$ and $\sigma$ are the location and shape parameters, respectively, and the corresponding probability density is
\begin{eqnarray}
g(x)=\frac{1}{\sigma}\exp\[ -\frac{x-\mu}{\sigma} -\exp\(\frac{x-\mu}{\sigma}\) \]
\end{eqnarray}

In Figure \ref{smax-gumbel}b are shown the probability densities of $s_{\text{max}}$ for $n=40, 60, 80, 100$ and $500$ events: independently on $n$, each density is in excellent agreement with the Gumbel distribution of extreme values, for the parameters $\mu= 1.743  \pm 0.002$ and $\sigma= 0.470   \pm 0.002$. Such values correspond to the mean and to the standard deviation of the distribution, $\tilde{\mu}\approx 2.00$ and $\tilde{\sigma}\approx 0.59$, respectively (see Appendix \ref{app-Gumbel}). It follow that the probability to obtain a maximum value of $s(\Theta)$, at any angular scale $\Theta$, greater or equal than a given value $\max\{s(\Theta)\}$ is
\begin{eqnarray}
p\(\max\{s(\Theta)\}\)=1-\exp\[ -\exp\(\frac{\max\{s(\Theta)\}-\mu}{\sigma}\) \],\nonumber
\end{eqnarray}
providing an analytical expression for the penalized probability defined in Eq. (\ref{def-p}).

\vspace{0.25truecm}\hspace{0.1truecm}\emph{Alternative hypothesis.} In order to investigate the behavior of MAF under the alternative hypothesis of an underlying anisotropic distribution of objects, we have generated anisotropic maps of $10^{4}$ skies, by varying the number of events from 20 to 100. In general, the anisotropy of a sky depends on several factors: for instance, in the case of cosmic rays, it depends on the distribution of sources, on magnetic fields and on propagation effects as energy loss or the GZK cutoff \cite{nagano2000observations, bhattacharjee2000origin} (and Ref. therein). Thus, a more complicated approach is required for the Monte Carlo realization of the maps. In order to estimate the power of MAF, we build reasonable anisotropic maps reflecting in part the real-world scenario, keeping in mind that our purpose is to build an anisotropic set of events for statistical analysis and not to generate events mimicking real data sets with the best available approximation. We proceed as follows: 
\begin{enumerate}
\item \emph{Catalog of candidate sources.} Although several models for production mechanisms of UHECR are available \cite{nagano2000observations, bhattacharjee2000origin} (and Ref. therein), \cite{hillas1984origin, hill1987ultra, berezinsky1997cosmic, berezinsky1997ultrahigh, venkatesan1997constraints, farrar1998correlation, fargion1999ultra, arons2003magnetars}, it is generally accepted that the candidate sources are extragalactic and trace the distribution of luminous matter on large scales \cite{waxman1997signature}. In particular, it has been shown that correlation with possible high redshift sources is unlikely \cite{sigl2001testing}, whereas compact sources are favored \cite{fodor2000ultrahigh, tinyakov2001correlation}: the recent result reported by the P. Auger Collaboration  experimentally supports the latter claim, showing a high correlation between the observed data and the distribution of nearby active galactic nuclei (AGN) \cite{auger2007correlation, auger2008correlation}. For these reasons, we use the Palermo Swift-BAT hard X-ray catalogue of AGN with known redshift within 200 Mpc ($z\leq0.047$) \cite{cusumano2010palermo}, as the reference catalog of candidate sources providing the most complete and uniform all-sky hard X-ray survey up to date.

\item \emph{Source effects.} Events, from each source in the reference catalog, are generated by weighting for the source flux and for the expected geometrical flux attenuation. Hence, the number of events coming from a source is proportional to its flux and to the factor $z^{-2}$: because of the small scales and the high energy of cosmic rays involved in anisotropy studies ($E\geq 4.0\times 10^{19}$ EeV), we assume a flat universe with zero cosmological constant ($\Omega=1$, $\Lambda=0$) and nonevolving source. Indeed, we naively take into account the possible deflections of the particles, due to the random component of the magnetic field, by producing arrival directions gaussianly-distributed with dispersion $\rho$ around the source. It is worth remarking that such a dispersion is strictly related to both the injection energy and the mass of the particle, as well as other physical quantities \cite{nagano2000observations}.

\item \emph{Background.} We take into account the possibility for a contaminating isotropic background of the anisotropy signal, by generating a number of events isotropically distributed, corresponding to a fraction $f_{\text{iso}}$ of the whole data set.

\item \emph{Detection effects.} As previously explained, the number of events detected by a single fully efficient and full-time operating surface detector, depends on its own relative exposure. In order to take into account such a detection effect, we generate the events according to the relative exposure of the single detector. Moreover, for each detector we generate the corresponding number of events reported in Table 1, in order to produce skies mimicking as much as possible real data currently available.
\end{enumerate}

However, for a more realistic distribution of events, several more constraints, in general based on further assumptions or debated models, are required: the mass of the particle, the injection spectrum of the source, the intervening magnetic field, to cite some of the most important. In our study of the MAF discrimination power, we fix $\rho=3^{\circ}$, as the mean angular deviation of UHECR in the galactic and extra-galactic magnetic field, and a background fraction $f_{\text{iso}}=0.3$.

\begin{figure}[!t]
	\centering
	  \includegraphics[width=14cm]{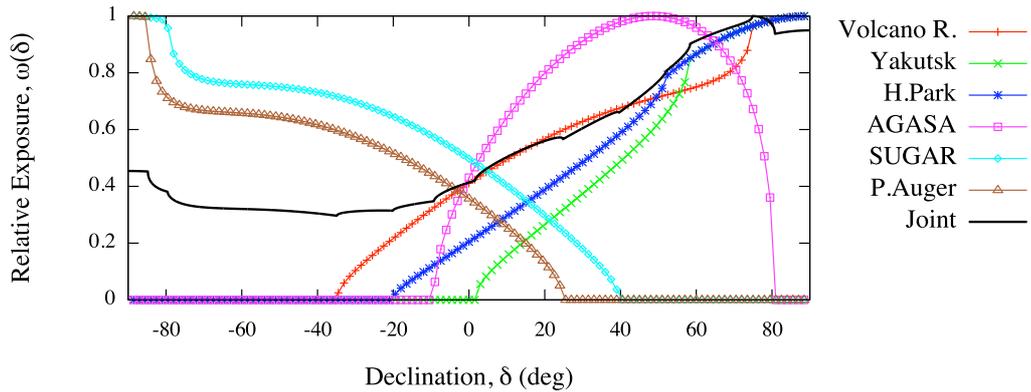}
	\caption{Relative geometrical exposure of each single detector listed in Table 1 (lines and points), and the joint exposure of all experiments (solid line).}
\label{all-expo}
\end{figure}

\begin{figure*}[!t]  
\centering
  	\subfigure[\, UHECR events and candidate sources.]{ 
	  \includegraphics[width=7.5cm]{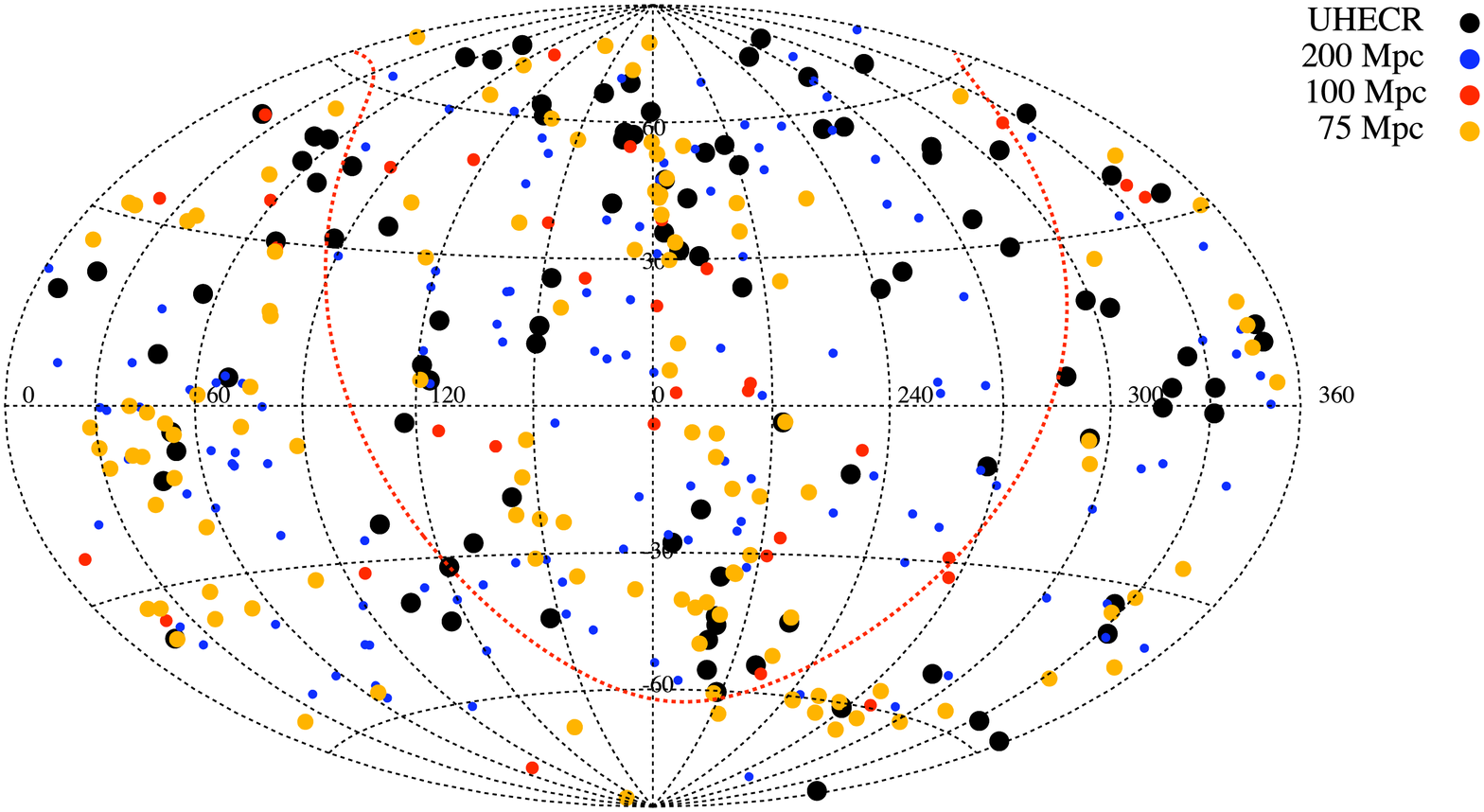}}
	\subfigure[\, Mock map.]{ 
	  \includegraphics[width=7.cm]{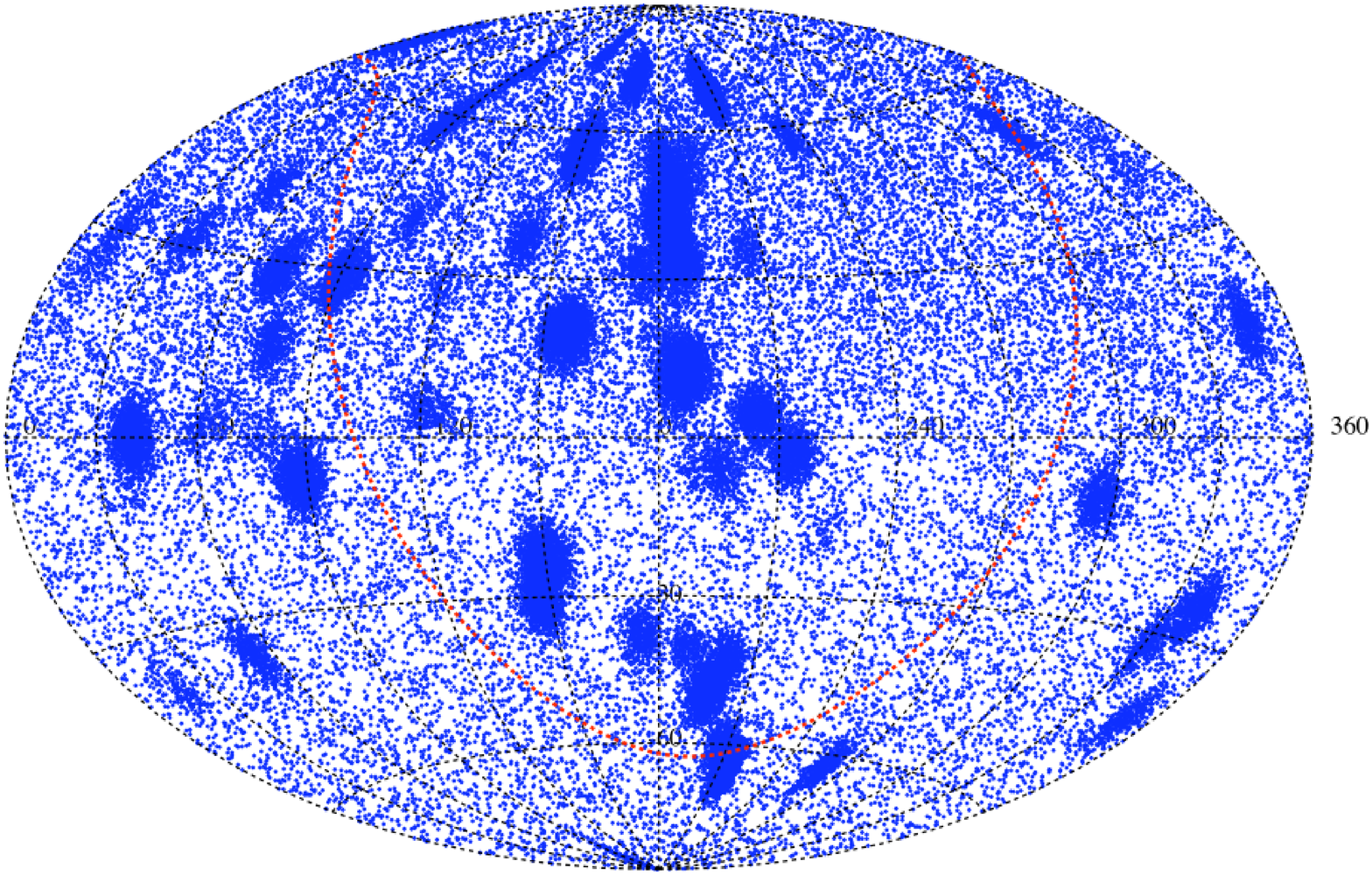}}
	  \caption{a) All-sky data set of 102 detected events with rescaled energy $E'\geq 40$ EeV (see the text for further information) superimposed on the distribution of AGN with known redshift ($z<0.047$) from the Palermo SWIFT-BAT hard X-ray catalogue; b) corresponding mock map generated for the statistical analysis (see details in the text). Equatorial coordinates are shown.}
\label{sky-data}
\end{figure*}

In order to produce a likely map of UHECR, we choose to generate events distributed in the whole sky, according to the number of events collected by surface detectors in the last decades. In particular, we consider events with energy $E\geq 4.0\times10^{19}$ EeV and error on the arrival direction smaller than $5^{\circ}$, as detected at the Sidney University Giant Airshower Recorder (SUGAR) \cite{winn1986arrival}, Akeno Giant Air Shower Array (AGASA) \cite{hayashida2000updated}, Haverah Park \cite{ave2000new}, Volcano Ranch (one event from \cite{nagano2000observations} and six events from \cite{medina2001ultra}), Yakutsk \cite{pravdin2005estimation}, P. Auger Observatory \cite{auger2008correlation}. However, the fluxes of particles as measured by those experiments do not agree each other in the absolute fluxes, and a rescaling is needed \cite{berezinsky2009ultra}. By assuming that the spectrum reported by the HiRes Collaboration \cite{abbasi2008first} corresponds to the correct energy scale, the rescaling, based on some specific characteristics of the UHECR spectrum, fixes the energy shift factors $\lambda$ for the other experiments \cite{berezinsky2009ultra, kachelriess2006clustering}. Positions, maximum zenith angles $\theta_{\text{max}}$, exposures and energy shift factors are reported in Table 1, for each experiment, as well as the number of detected events with rescaled energy $E'\geq 4.0\times10^{19}$ EeV ($E'=\lambda E$). In Figure \ref{all-expo} is shown the relative geometrical exposure of each single detector listed in Table 1, as well as the joint exposure of all experiments. For reference, in Figure \ref{sky-data}a is shown the all-sky data set of 102 detected events with rescaled energy $E'$, superimposed on the distribution of AGN within 200 Mpc from the reference catalog, whereas in Figure \ref{sky-data}b is shown the mock map of simulated events according to physical constraints previously described.

\begin{table}[!t]
\centering
\begin{tabular}{lcllcc}
\hline
\hline
\textbf{Experiment} & $\phi_{0}$ & $\theta_{\text{max}}$ & \textbf{Exp.} ($m^{2}\text{ s sr}$) & $\lambda$ & \#\textbf{Ev.}\\
\hline
\hline
Volcano R. & $35.15^{\circ}$N  &  $70^{\circ}$ & $0.2\times 10^{16}$ & 1.000 & 6 \\
Yakutsk & $61.60^{\circ}$N &  $60^{\circ}$       & $1.8\times 10^{16}$ & 0.625 & 20   \\
H. Park & $53.97^{\circ}$N &  $74^{\circ}$       & $-$                            & 1.000 & 7    \\
AGASA   & $35.78^{\circ}$N &  $45^{\circ}$     & $4.0\times 10^{16}$ & 0.750 & 29   \\
SUGAR   & $30.43^{\circ}$S & $70^{\circ}$      & $5.3\times 10^{16}$ & 0.500 & 13   \\
P. Auger & $35.20^{\circ}$S &  $60^{\circ}$      & $28.4\times 10^{16}$ & 1.200 & 27   \\
\hline
\hline
\end{tabular}
\caption{Surface detectors: positions, maximum zenith angles $\theta_{\text{max}}$, exposures, energy shift factors and number of detected events with rescaled energy $E'\geq 4.0\times10^{19}$ EeV.}
\end{table}

\begin{figure}[!h]
	\centering
	  \includegraphics[width=12cm]{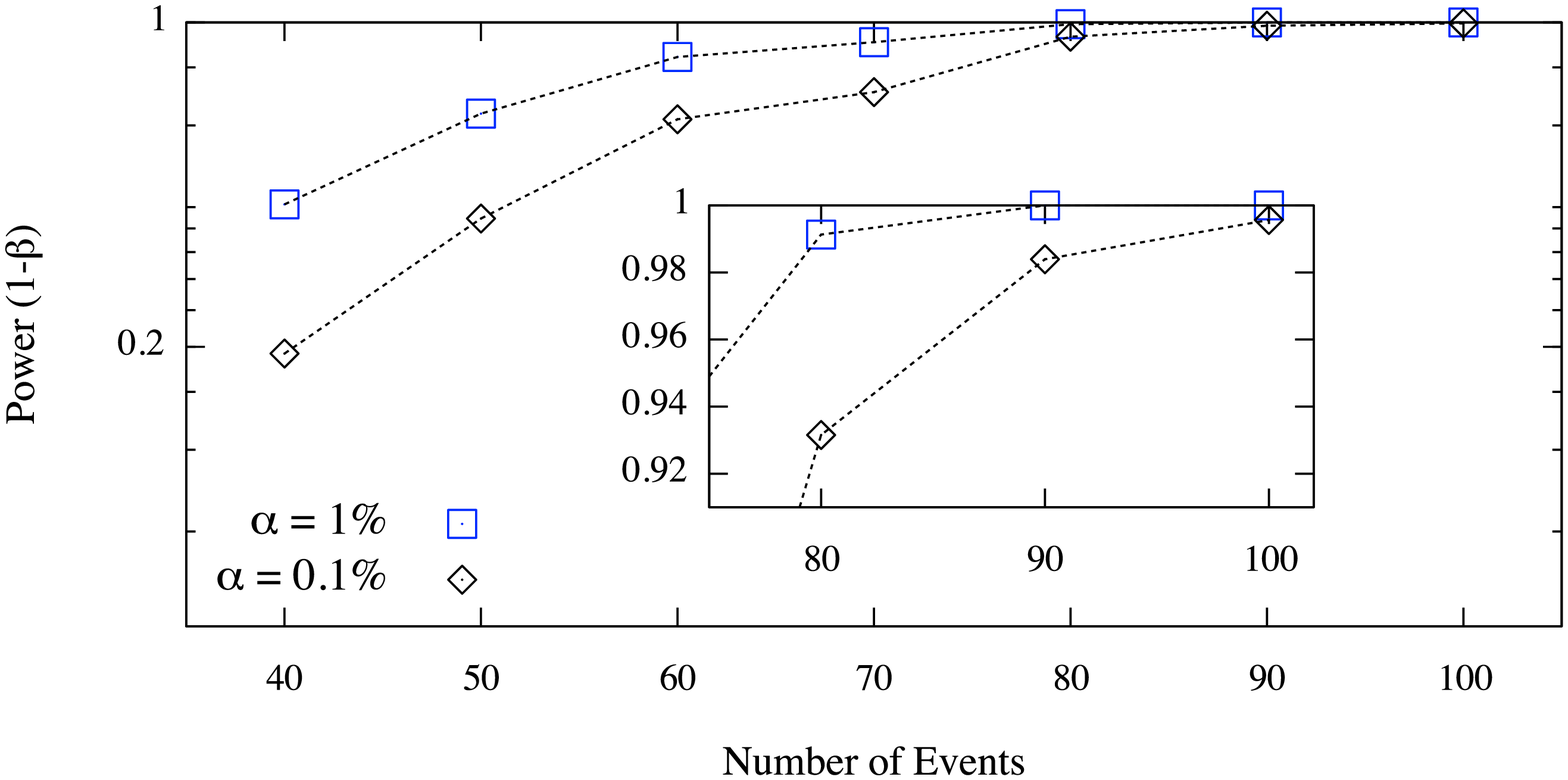}
	\caption{MAF power vs. the number of events sampled from anisotropic mock maps generated as described in the text, for values of the significance corresponding to $\alpha=0.1\%$ and $\alpha=1\%$.}
\label{maf-power}
\end{figure}

In Figure \ref{maf-power} we show the power $1-\beta$ vs. the number of events, generated as described above. A sky is labelled as \emph{anisotropic} if, for a fixed value of the significance $\alpha$, the penalized chance probability as defined in Eq. (\ref{def-p}) is lesser or equal than $\alpha$, i.e. if the condition 
\begin{eqnarray}
\tilde{p}(\Theta^{\star})=\arg\min_{\Theta} p(\Theta)\leq \alpha\nonumber
\end{eqnarray}
holds for some angular scale $\Theta^{\star}$. In Figure \ref{maf-power} is shown the power for two values of the significance threshold, namely $\alpha=0.1\%$ and $\alpha=1\%$, estimated through the analytical approach. For applications, a power of 90\% is generally required: under this threshold the method could miss to detect an existing anisotropy signal. In the case of the MAF, and for the considered anisotropic mock map, the power increases with the number of events $n$ and it is able to detect the anisotropic signal for $n\geq 60$, with significance $\alpha=1\%$. However, by decreasing the significance for the statistical test, the power requires a greater number of events to reach the 90\% threshold, as expected: our test clearly shows that the MAF provides an excellent discrimination power for $n\geq80$. Indeed, we verified the agreement between the analytical and the Monte Carlo estimations of the discrimination power.



\section{Discussion and conclusion}

We introduced a new statistical test, based on a multiscale approach, for detecting an anisotropy signal in the arrival direction distribution of UHECR, that makes use of an information theoretical measure of similarity, namely the Kullback-Leibler divergence, and of the extreme value theory. Within the present work we showed that our procedure is suitable for the analysis of both small and large data sets of events, by applying it on several Monte Carlo realizations of isotropic and anisotropic synthetic data sets, corresponding to plausible scenarios in the physics of highest energy cosmic rays. In fact, for small data sets as well as for larger ones, the method is able to recover the information about the most significant angular scale of clustering in the data, even in presence of strong isotropic contamination. 

The advantages of our approach over other methods are multiples. First, the method allows an analytical description of quantities involved in the estimation of the amount of anisotropy signal in the data, avoiding thousands of Monte Carlo realizations needed for the penalizing procedure of results and drastically reducing the computation time. Second, the method allows the detection of a physical observable, namely the clustering scale, in the case of a point source. In the case of multiple sources, the information is about the most significant clustering scale(s), according to source distribution. Third, the method is unbiased against the null hypothesis and it provides a high discrimination power even in presence of strong contaminating isotropic background, for both small and large data sets. Although in this work we referred to UHECR physics for our applications, it is worth remarking that the method is suitable for the detection of the anisotropy signal in each data set involving a distribution of angular coordinates on the sphere, and it can be adapted to non-spherical spaces by properly redefining the dynamical counting algorithm.

 \acknowledgments
 Authors thank the P. Auger Collaboration for generous comments and fruitful discussions, and, in particular, O. Deligny for his unvaluable suggestions during the definition of our procedure and P.L. Ghia for strongly encouraging this work. Another special thank is for R. Sinatra for the useful discussions and for the anonymous referee, for the useful comments and suggestions. Finally, we thank the ``Fondo per il potenziamento per la ricerca in informatica'', (Dipartimento di Matematica e Informatica, Università degli Studi di Palermo, Palermo, Italy) for having kindly made available the computing resources required for the analyses presented within the present work.


\appendix
\section{The Kullback-Leibler divergence}\label{app-KL}

Let $P$ and $Q$ be two probability distributions, with densities $p(x)$ and $q(x)$, respectively. The Kullback-Leibler (KL) divergence is a measure quantifying the error in approximating the density $p(x)$ by means of $q(x)$, and it is defined \cite{kullback1951information, kullback1987kullback} as
\begin{eqnarray}
\mathcal{D}_{KL}(p||q)=\int p(x)\log\frac{p(x)}{q(x)}dx
\end{eqnarray}
The KL divergence is non-negative, i.e. $\mathcal{D}_{KL}(p||q)\geq 0$ with equality if and only if $P=Q$, and asymmetric, i.e. $\mathcal{D}_{KL}(p||q)\neq \mathcal{D}_{KL}(q||p)$. The statistical interpretation of KL divergence is as follows.

Let $\tilde{P}$ the empirical distribution of random outcomes $x_{i}$ ($i=1,2,...,n$) of the true distribution $P$, putting the probability $\frac{1}{n}$ on each outcome as
\begin{eqnarray}
\label{empdistrib}
\tilde{p}(x)=\frac{1}{n}\sum_{i=1}^{n}\delta(x-x_{i})
\end{eqnarray}
and let $Q_{\Theta}$ be the statistical model for the data, depending on the unknown parameter $\Theta$. It follows
\begin{eqnarray}
\label{kldiv}\mathcal{D}_{KL}(\tilde{p}||q_{\Theta})=-\mathcal{H}(\tilde{p})-\int\tilde{p}(x)\log q(x|\Theta)dx
\end{eqnarray}
where $\mathcal{H}(\tilde{p})$ is the information entropy of $\tilde{p}$, not depending on $\Theta$, whereas $\tilde{p}$ and $q_{\Theta}=q(x|\Theta)$ are the corresponding densities of $\tilde{P}$ and $Q_{\Theta}$, respectively. Putting Eq. (\ref{empdistrib}) in the right-hand side of Eq. (\ref{kldiv}):
\begin{eqnarray}
\mathcal{D}_{KL}(\tilde{p}||q_{\Theta})&=&-\mathcal{H}(\tilde{p})-\frac{1}{n}\sum_{i=1}^{n}\log q(x_{i}|\Theta)\nonumber\\
&=&-\mathcal{H}(\tilde{p})-\frac{1}{n}\mathcal{L}_{q}(\Theta | x)
\end{eqnarray}
where $\mathcal{L}_{q}(\Theta | x)$ is the log-likelihood of the statistical model. It directly follows that
\begin{eqnarray}
\arg\min_{\Theta}\mathcal{D}_{KL}(\tilde{p}||q_{\Theta})=\frac{1}{n}\arg\max_{\Theta}\mathcal{L}_{q}(\Theta | x)
\end{eqnarray}
where the function $\arg\min (\arg\max) f(\Theta)$ retrieves the minimum (maximum) of the function $f(\Theta)$. Hence, another way to obtain the maximum likelihood estimation it to minimize the KL divergence \cite{cover1991elements}; indeed, it can be shown that the KL divergence corresponds to the expected log-likelihood ratio \cite{eguchi2006interpreting}.

\section{The Gumbel distribution}\label{app-Gumbel}

Extreme value theory is the research area dealing with the statistical analysis of the extremal values of a stochastic variable. Let $x_{i}$ ($i=1,2,...,n$) be i.i.d. random outcomes of a distribution $F$. If $M_{n}=\max\{x_{1},x_{2},...,x_{n}\}$, the probability to obtain an outcome greater or equal than $M_{n}$ is:
\begin{eqnarray}
\text{Pr}(M_{n}\leq x)=\text{Pr}(x_{1}\leq x, x_{2}\leq x, ..., x_{n}\leq x)= F^{n}(x)\nonumber
\end{eqnarray}
It can be shown that the limiting distribution $F^{n}(x)$ is degenerate and should be normalized \cite{dehaan2006extreme}. However, if there exists sequences of real constants $a_{n}>0$ and $b_{n}$ such that
\begin{eqnarray}
\text{Pr}\(\frac{M_{n}-b_{n}}{a_{n}}\leq x\)= F^{n}(a_{n}x+b_{n})\nonumber
\end{eqnarray}
then 
\begin{eqnarray}
\lim_{n\lto \infty}F^{n}(a_{n}x+b_{n})=G(x)
\end{eqnarray}
The function $G(x)$ is the generalized extreme value (GEV) or Fisher-Tippett distribution
\begin{eqnarray}
G(z)=\left\{
\begin{array}{ll}
\exp\(-e^{-z}\) & \xi=0\\
\exp\[ -\(1-\xi z\)^{\frac{1}{\xi}} \] & \xi\neq 0
\end{array}
\right.,\quad z=\frac{x-\mu}{\sigma}
\end{eqnarray}
defined for $1-\xi z>0$ if $\xi\neq 0$ and for $z\in\mathbb{R}$ if $\xi=0$, where $\mu, \sigma$ and $\xi$ are the location, scale and shape parameters, respectively. The Gumbel distribution is related to the distribution of maxima \cite{gumbel1954statistical,gumbel2004statistics} and it is retrieved for $\xi=0$ \cite{dehaan2006extreme}. The corresponding probability density $g(x)$ is easily obtained from $G$ as
\begin{eqnarray}
\label{gumbeldef}g(x)=\frac{1}{\sigma}\exp\[ -\frac{x-\mu}{\sigma} -\exp\(\frac{x-\mu}{\sigma}\) \]
\end{eqnarray}
Finally, the two parameters $\mu$ and $\sigma$ can be related to the mean $\tilde{\mu}$ and to the standard deviation $\tilde{\sigma}$ of the distribution, by means of the following relations:
\begin{eqnarray}
\tilde{\mu}&=& \mu + \gamma \sigma\\
\tilde{\sigma}^{2}&=& \frac{\pi^{2}}{6}\sigma^{2}
\end{eqnarray}
where $\gamma=0.577215...$ is the Euler constant.

\newpage
\addcontentsline{toc}{section}{References} 
\begin{small}
\bibliographystyle{JHEP} 
\bibliography{draft}

\providecommand{\href}[2]{#2}\begingroup\raggedright\begin{thebibliography}{10}

\bibitem{Davis-1983}
{M. Davis and P.J.E. Peebles}, {\it A survey of galaxy redshifts vs the two
  point position and velocity correlations},  {\em Ap. J.} {\bf 267} (1983)
  465--482.

\bibitem{Szalay-1993}
{S.D. Landy and A.S. Szalay}, {\it Bias and variance of angular correlation
  functions},  {\em Ap. J.} {\bf 412} (1993) 64--71.

\bibitem{Hamilton-1993}
{A.J.S. Hamilton}, {\it Toward better ways to measure the galaxy correlation
  function},  {\em Ap. J.} {\bf 417} (1993) 19--35.

\bibitem{Peebles-1980}
{P.J.E. Peebles}, {\em The Large Scale Structure of the Universe}.
\newblock Princeton Univ. Press, 1980.

\bibitem{cuoco2006first}
A.~Cuoco, G.~Miele, and P.~Serpico, {\it {First hints of large scale structures
  in the ultrahigh energy sky?}},  {\em Phys. Rev. D} {\bf 74} (2006), no.~12
  123008.

\bibitem{cuoco2008clustering}
A.~Cuoco, S.~Hannestad, T.~Haugb{\o}lle, M.~Kachelrie{\ss}, and P.~Serpico,
  {\it {Clustering Properties of Ultra-High-Energy Cosmic Rays}},  {\em The
  Astrophysical Journal} {\bf 676} (2008) 807--815.

\bibitem{cuoco2009global}
A.~Cuoco, S.~Hannestad, T.~Haugb{\o}lle, M.~Kachelrie{\ss}, and P.~Serpico,
  {\it {A global autocorrelation study after the first Auger data}},  {\em
  Astrop. J.} {\bf 702} (2009) 825--832.

\bibitem{kachelriess2005ultra}
M.~Kachelrie{\ss} and D.~Semikoz, {\it {Ultra-high energy cosmic rays from a
  finite number of point sources}},  {\em Astrop. Phys.} {\bf 23} (2005), no.~5
  486--492.

\bibitem{kachelriess2006clustering}
M.~Kachelrie{\ss} and D.~Semikoz, {\it {Clustering of ultra-high energy cosmic
  ray arrival directions on medium scales}},  {\em Astrop. Phys.} {\bf 26}
  (2006), no.~1 10--15.

\bibitem{ave20092pt+}
M.~Ave, L.~Cazon, J.~Cronin, J.~de~Mello~Neto, A.~Olinto, V.~Pavlidou,
  P.~Privitera, B.~Siffert, F.~Schmidt, and T.~Venters, {\it {The 2pt+: an
  enhanced 2 point correlation function}},  {\em J. Cosm. Astrop. Phys.} {\bf
  2009} (2009) 023.

\bibitem{Hague09}
{J.D. Hague, B.R. Becker, M.S. Gold and J.A.J. Matthews}, {\it A three-point
  cosmic ray anisotropy method},  {\em J. Phys. G: Nucl. Part. Phys.} {\bf 36}
  (2009) 115203.

\bibitem{VCV06}
{M.P. Veron-Cetty and P. Veron}, {\it Quasars and active galactic nuclei (12th
  ed.)},  {\em Astron. Astrop.} {\bf 455} (2006), no.~2 773.

\bibitem{MelloNeto2009}
{J. de Mello Neto \it{et al}}, {\it {Search for intrinsic anisotropy in the
  UHECRs data from the Pierre Auger Observatory}},  {\em Proc. 31st ICRC, Lodz}
  (2009).

\bibitem{stokes2004using}
B.~Stokes, C.~Jui, and J.~Matthews, {\it {Using fractal dimensionality in the
  search for source models of ultra-high energy cosmic rays}},  {\em Astrop.
  Phys.} {\bf 21} (2004), no.~1 95--109.

\bibitem{kullback1951information}
S.~Kullback and R.~Leibler, {\it {On information and sufficiency}},  {\em The
  Annals of Mathematical Statistics} {\bf 22} (1951) 79--86.

\bibitem{kullback1987kullback}
S.~Kullback, {\it {The Kullback-Leibler distance}},  {\em The American
  Statistician} {\bf 41} (1987) 340--341.

\bibitem{akaike1973information}
H.~Akaike, {\it {Information theory and an extension of the maximum likelihood
  principle}},  in {\em 2nd Int. Symp. on Inform. Th.}, pp.~267--281, 1973.

\bibitem{akaike1974new}
H.~Akaike, {\it {A new look at the statistical model identification}},  {\em
  IEEE trans. on autom. control} {\bf 19} (1974), no.~6 716--723.

\bibitem{anderson2000null}
D.~Anderson, K.~Burnham, and W.~Thompson, {\it {Null hypothesis testing:
  problems, prevalence, and an alternative}},  {\em J. Wil. manag.} {\bf 64}
  (2000), no.~4 912--923.

\bibitem{plastino1997relationship}
A.~Plastino, A.~Plastino, and H.~Miller, {\it {On the relationship between the
  Fisher-Frieden-Soffer arrow of time, and the behaviour of the Boltzmann and
  Kullback entropies}},  {\em Phys. Lett. A} {\bf 235} (1997), no.~2 129--134.

\bibitem{plastino1997minimum}
A.~Plastino, H.~Miller, and A.~Plastino, {\it {Minimum Kullback entropy
  approach to the Fokker-Planck equation}},  {\em Phys. Rev. E} {\bf 56}
  (1997), no.~4 3927--3934.

\bibitem{portesi2007geometrical}
M.~Portesi, F.~Pennini, and A.~Plastino, {\it {Geometrical aspects of a
  generalized statistical mechanics}},  {\em Physica A} {\bf 373} (2007)
  273--282.

\bibitem{fuchs1995distinguishability}
C.~Fuchs, {\it {Distinguishability and accessible information in quantum
  theory}},  \href{http://xxx.lanl.gov/abs/quant-ph/9601020v1}{{\tt
  quant-ph/9601020v1}}.

\bibitem{reginatto1998derivation}
M.~Reginatto, {\it {Derivation of the equations of nonrelativistic quantum
  mechanics using the principle of minimum Fisher information}},  {\em Phys.
  Rev. A} {\bf 58} (1998), no.~3 1775--1778.

\bibitem{abe1999quantum}
S.~Abe and A.~Rajagopal, {\it {Quantum entanglement inferred by the principle
  of maximum nonadditive entropy}},  {\em Phys. Rev. A} {\bf 60} (1999), no.~5
  3461--3466.

\bibitem{abe2003nonadditive}
S.~Abe, {\it {Nonadditive generalization of the quantum Kullback-Leibler
  divergence for measuring the degree of purification}},  {\em Phys. Rev. A}
  {\bf 68} (2003), no.~3 32302.

\bibitem{gersch1979automatic}
W.~Gersch, F.~Martinelli, J.~Yonemoto, M.~Low, and J.~Mc~Ewan, {\it {Automatic
  classification of electroencephalograms: Kullback-Leibler nearest neighbor
  rules}},  {\em Science} {\bf 205} (1979), no.~4402 193.

\bibitem{burnham2001kullback}
K.~Burnham and D.~Anderson, {\it {Kullback-Leibler information as a basis for
  strong inference in ecological studies}},  {\em Wil. Res.} {\bf 28} (2001),
  no.~2 111--120.

\bibitem{sommers2001cosmic}
P.~Sommers, {\it {Cosmic ray anisotropy analysis with a full-sky observatory}},
   {\em Astrop. Phys.} {\bf 14} (2001), no.~4 271--286.

\bibitem{gumbel1954statistical}
E.~Gumbel, {\em {Statistical theory of extreme values and some practical
  applications: A series of lectures}}.
\newblock National Bureau of Standards Washington DC, 1954.

\bibitem{gumbel2004statistics}
E.~Gumbel, {\em {Statistics of extremes}}.
\newblock Dover Pub., 2004.

\bibitem{nagano2000observations}
M.~Nagano and A.~Watson, {\it {Observations and implications of the
  ultrahigh-energy cosmic rays}},  {\em Rev. Mod. Phys.} {\bf 72} (2000), no.~3
  689--732.

\bibitem{bhattacharjee2000origin}
P.~Bhattacharjee and G.~Sigl, {\it {Origin and propagation of extremely
  high-energy cosmic rays}},  {\em Phys. Rep.} {\bf 327} (2000), no.~3-4
  109--247.

\bibitem{hillas1984origin}
A.~Hillas, {\it {The origin of ultra-high-energy cosmic rays}},  {\em Ann. Rev.
  Astr. Astrop.} {\bf 22} (1984), no.~1 425--444.

\bibitem{hill1987ultra}
C.~Hill, D.~Schramm, and T.~Walker, {\it {Ultra-high-energy cosmic rays from
  superconducting cosmic strings}},  {\em Phys. Rev. D} {\bf 36} (1987), no.~4
  1007--1016.

\bibitem{berezinsky1997cosmic}
V.~Berezinsky and A.~Vilenkin, {\it {Cosmic necklaces and ultrahigh energy
  cosmic rays}},  {\em Phys. Rev. Lett.} {\bf 79} (1997), no.~26 5202--5205.

\bibitem{berezinsky1997ultrahigh}
V.~Berezinsky, M.~Kachelrie{\ss}, and A.~Vilenkin, {\it {Ultrahigh Energy
  Cosmic Rays without Greisen-Zatsepin-Kuzmin Cutoff}},  {\em Phys. Rev. Lett.}
  {\bf 79} (1997), no.~22 4302--4305.

\bibitem{venkatesan1997constraints}
A.~Venkatesan, M.~Miller, and A.~Olinto, {\it {Constraints on the production of
  ultra-high-energy cosmic rays by isolated neutron stars}},  {\em Ap. J.} {\bf
  484} (1997) 323.

\bibitem{farrar1998correlation}
G.~Farrar and P.~Biermann, {\it {Correlation between compact radio quasars and
  ultrahigh energy cosmic rays}},  {\em Phys. Rev. Lett.} {\bf 81} (1998),
  no.~17 3579--3582.

\bibitem{fargion1999ultra}
D.~Fargion, B.~Mele, and A.~Salis, {\it {Ultra-High-Energy Neutrino Scattering
  onto Relic Light Neutrinos in the Galactic Halo as a Possible Source of the
  Highest Energy Extragalactic Cosmic Rays}},  {\em Ap. J.} {\bf 517} (1999)
  725.

\bibitem{arons2003magnetars}
J.~Arons, {\it {Magnetars in the metagalaxy: an origin for ultra-high-energy
  cosmic rays in the nearby universe}},  {\em Ap. J.} {\bf 589} (2003) 871.

\bibitem{waxman1997signature}
E.~Waxman, K.~Fisher, and T.~Piran, {\it {The Signature of a Correlation
  between Cosmic-Ray Sources above 1019 eV and Large-Scale Structure}},  {\em
  Ap. J.} {\bf 483} (1997) 1.

\bibitem{sigl2001testing}
G.~Sigl, D.~Torres, L.~Anchordoqui, and G.~Romero, {\it {Testing the
  correlation of ultrahigh energy cosmic rays with high redshift sources}},
  {\em Phys. Rev. D} {\bf 63} (2001), no.~8 81302.

\bibitem{fodor2000ultrahigh}
Z.~Fodor and S.~Katz, {\it {Ultrahigh energy cosmic rays from compact
  sources}},  {\em Phys. Rev. D} {\bf 63} (2000), no.~2 23002.

\bibitem{tinyakov2001correlation}
P.~Tinyakov and I.~Tkachev, {\it {Correlation function of ultrahigh-energy
  cosmic rays favors point sources}},  {\em JETP Lett.} {\bf 74} (2001), no.~1
  1--5.

\bibitem{auger2007correlation}
{J. Abraham \it{et al}}, {\it {Correlation of the Highest-Energy Cosmic Rays
  with Nearby Extragalactic Objects}},  {\em Science} {\bf 318} (2007),
  no.~5852 938.

\bibitem{auger2008correlation}
{J. Abraham \it{et al}}, {\it {Correlation of the highest-energy cosmic rays
  with the positions of nearby active galactic nuclei}},  {\em Astrop. Phys.}
  {\bf 29} (2008), no.~3 188--204.

\bibitem{cusumano2010palermo}
{G. Cusumano \it{et al}}, {\it {The Palermo Swift-BAT hard X-ray catalogue}},
  {\em Astron. Astrop.} {\bf 510} (2010).

\bibitem{winn1986arrival}
M.~Winn, J.~Ulrichs, L.~Peak, C.~McCusker, and L.~Horton, {\it {The arrival
  directions of cosmic rays above $10^{17}$ eV}},  {\em J. Phys. G: Nuc. Phys.}
  {\bf 12} (1986) 675.

\bibitem{hayashida2000updated}
{N. Hayashida \it{et al}}, {\it {Updated AGASA event list above
  $4\times10^{19}eV$}},  \href{http://xxx.lanl.gov/abs/astro-ph/0008102}{{\tt
  astro-ph/0008102}}.

\bibitem{ave2000new}
M.~Ave, J.~Hinton, R.~Vazquez, A.~Watson, and E.~Zas, {\it {New Constraints
  from Haverah Park Data on the Photon and Iron Fluxes of Ultrahigh-Energy
  Cosmic Rays}},  {\em Phys. Rev. Lett.} {\bf 85} (2000), no.~11 2244--2247.

\bibitem{medina2001ultra}
G.~Medina-Tanco, {\it {Ultra-high energy cosmic rays: are they isotropic?}},
  {\em Ap. J.} {\bf 549} (2001) 711.

\bibitem{pravdin2005estimation}
{Pravdin, M.I. \it{et al}}, {\it {Estimation of the giant shower energy at the
  Yakutsk EAS Array}},  {\em Proc. 29th ICRC, Pune} {\bf 7} (2005) 243--246.

\bibitem{berezinsky2009ultra}
V.~Berezinsky, {\it {Ultra High Energy Cosmic Ray Protons: Signatures and
  Observations}},  {\em Nucl. Phys. B - Proc. Supp.} {\bf 188} (2009) 227--232.

\bibitem{abbasi2008first}
{R.U. Abbasi \it{et al}}, {\it {First observation of the
  Greisen-Zatsepin-Kuzmin suppression}},  {\em Phys. Rev. Lett.} {\bf 100}
  (2008), no.~10 101101.

\bibitem{cover1991elements}
T.~Cover and J.~Thomas, {\em {Elements of information theory}}.
\newblock Wiley Series in Telecommunications, 1991.

\bibitem{eguchi2006interpreting}
S.~Eguchi and J.~Copas, {\it {Interpreting kullback-leibler divergence with the
  neyman-pearson lemma}},  {\em J. Multiv. Anal.} {\bf 97} (2006), no.~9
  2034--2040.

\bibitem{dehaan2006extreme}
L.~De~Haan and A.~Ferreira, {\em {Extreme value theory: an introduction}}.
\newblock Springer Verlag, 2006.

\end{thebibliography}\endgroup
\end{small}

\end{document}